\documentclass[10pt,twocolumn,showpacs,pre,preprintnumbers,amsmath,amssymb,aps]{revtex4-1}
\usepackage{graphicx,dcolumn,bm,color,comment,subfigure,upgreek}
\newcommand{\unit}[1]{\ensuremath{\, \mathrm{#1}}} 
\renewcommand{\>}{\rangle} \newcommand{\ka}{K_A}
\newcommand{\kb}{k_\textrm{\tiny B}}
\newcommand{\kon}{k_\textrm{\scriptsize{on}}}
\newcommand{\koff}{k_\textrm{\scriptsize{off}}}

\newcommand{\ton}{\cT_\textrm{\scriptsize{on}}}
\newcommand{\toff}{\cT_\textrm{\scriptsize{off}}}

 \newcommand{\cL}{\mathcal{L}}
\newcommand{\cT}{\mathcal{T}} 
\newcommand{\heff}{h_\textrm{\scriptsize eff}}

\newcommand{\qmle}{\hat{q}_\textrm{\tiny{MLE}}}
\newcommand{\Qmle}{\hat{Q}_\textrm{\tiny{MLE}}}
\newcommand{\phimle}{\hat{\phi}_\textrm{\tiny{MLE}}}
 \newcommand{\btheta}{\mathbf{\Theta}}
\begin{document}
\preprint{Phys. Rev. E}
\title{Physical Limits on Cellular Directional Mechanosensing}
\author{Roland Bouffanais$^{1,2}$}
\author{Jianmin Sun$^{1}$}
\author{Dick K. P. Yue$^2$}
\affiliation{$^1$Singapore University of Technology and Design, 20 Dover
  Drive, Singapore 138682} \affiliation{$^2$Department of Mechanical
  Engineering, Massachusetts Institute of Technology, Cambridge, Massachusetts
  02139, USA }%
\date{\today}
\begin{abstract}
  Many eukaryotic cells are able to perform directional mechanosensing by
  directly measuring minute spatial differences in the mechanical stress on
  their membranes. Here, we explore the limits of a single mechanosensitive
  channel activation using a two-state double-well model for the gating
  mechanism. We then focus on the physical limits of directional
  mechanosensing by a single cell having multiple mechanosensors and subjected
  to a shear flow inducing a nonuniform membrane tension. Our results
  demonstrate that the accuracy in sensing the mechanostimulus direction not
  only increases with cell size and exposure to a signal, but also grows for
  cells with a near-critical membrane prestress. Finally, the existence of a
  nonlinear threshold effect, fundamentally limiting the cell's ability to
  effectively perform directional mechanosensing at a low signal-to-noise
  ratio, is uncovered.
\end{abstract}
\maketitle
\section{Introduction}
% -----------------------------

%
Cells usually dwell in complex microenvironments and, therefore, are
inherently sensitive to a variety of biomechanical stimuli, such as blood flow
and organ distensions, which induce mechanical stresses in the membrane and
cytoskeleton of cells. Recent studies indicate that mechanical forces have a
far greater impact on cell functions than previously appreciated. Eukaryotic
cells, such as epithelial cells, amoebae, and neutrophils, are remarkably
sensitive to shear flow direction~\cite{Arnadottir,ShearFlow,Park,Moares}.

More quantitatively, endothelial cells have been found to respond to laminar
shear stress levels in the range of 0.02--0.16~Pa with a cellular alignment in
the direction of the flow for a shear stress beyond
0.5~Pa~\cite{Olesen,Davies}. In other instances, some eukaryotic cells
performed parallel or perpendicular cellular alignment to the shear flow
direction~\cite{Park}. \textit{Xenopus laevis} oocytes were found to respond
to laminar shear stress of magnitude 0.073~Pa, whereas, the amoeba
\textit{Dictyostelium discoideum} exhibits shear-flow induced motility in the
direction of creeping flows with shear stresses as low as
0.7~Pa~\cite{Decave}. Similar magnitudes of this shear-stress based
mechanostimulus for other types of cells are reported in Ref.~\cite{Orr}. To
better appreciate the exquisite sensitivity of those cells~\cite{S&S}, it is
worth highlighting the minuteness of those mechanostimuli. For instance, a
characteristic shear stress of magnitude $\sigma \sim 1$~Pa generates a
maximum excess membrane tension $\Delta \gamma_{\textrm{\tiny max}}\sim \sigma
R$, which for a typical cell size of $R\sim 10~\upmu\textrm{m}$ is on the
order of $10~\upmu\textrm{N}\cdot\textrm{m}^{-1}$. According to Rawicz
\textit{et al.}~\cite{Rawicz}, such a value represents a minuscule membrane
tension. Furthermore, this membrane tension induced by the shear stress is 1
or 2 orders of magnitude smaller than typical lytic resting membrane tensions:
$\gamma_0\sim 1$--$2~\textrm{mN}\cdot\textrm{m}^{-1}$~\cite{Opsahl}.  From the
dynamical standpoint, the lower the shear rate, the longer the exposure
required for a cell to respond~\cite{Moares}. Finally, mechanosensing has been
shown to be of paramount importance to self-organizing behaviors of those
social cells~\cite{Bouffanais}.

% ---------------------------------------
\begin{figure}[htbp]
  \subfigure[] {
    \includegraphics[width=3.6cm]{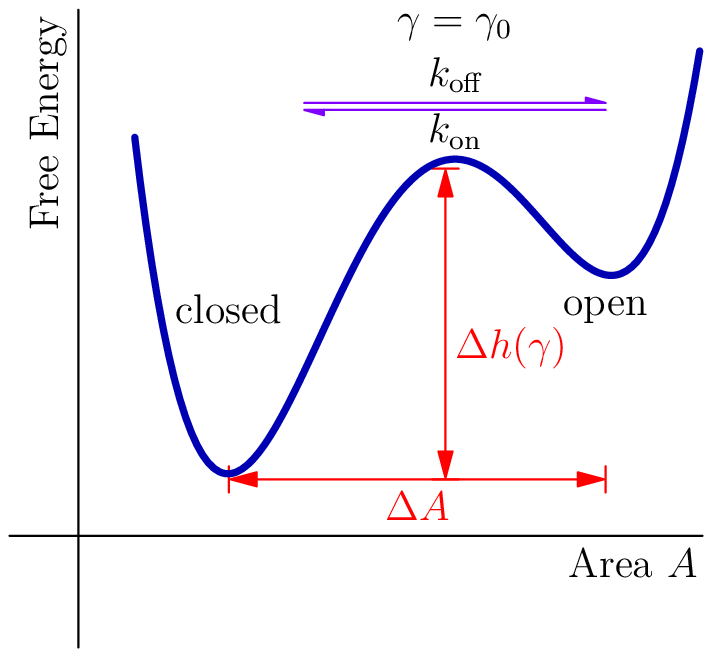}
    \label{Figure1a}} \subfigure[] {
    \includegraphics[width=3.6cm]{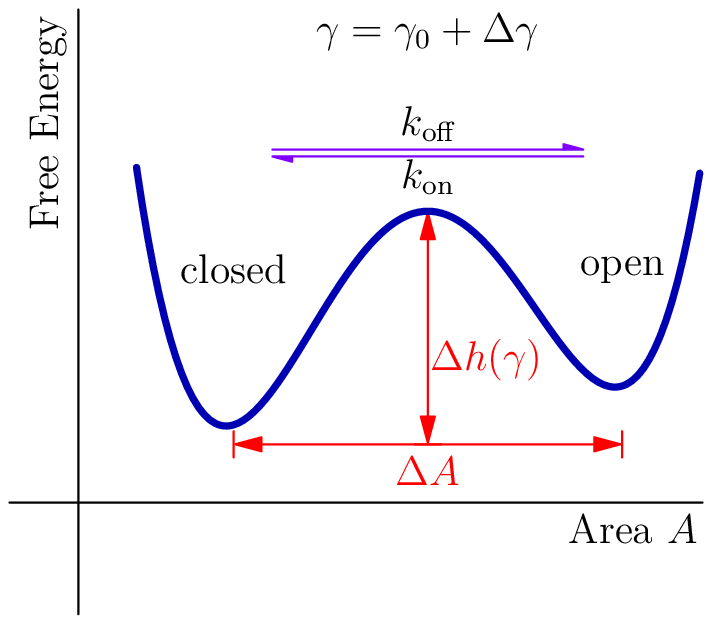}
    \label{Figure1b}} \subfigure[] {
    \includegraphics[width=3.6cm]{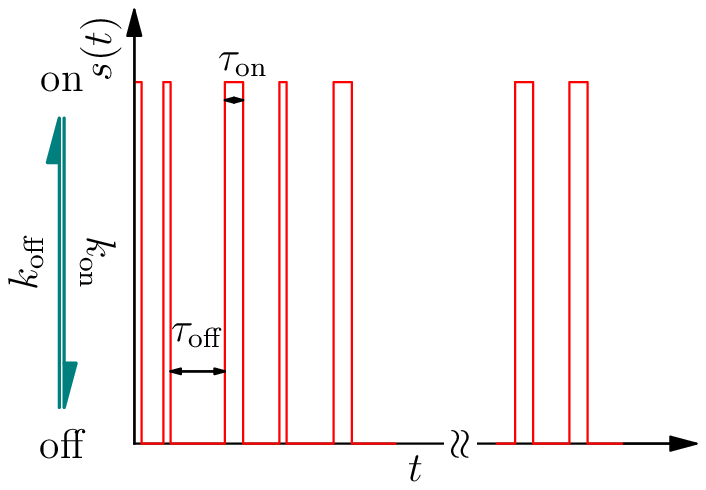}
    \label{Figure1c}} \subfigure[] {
    \includegraphics[width=3.6cm]{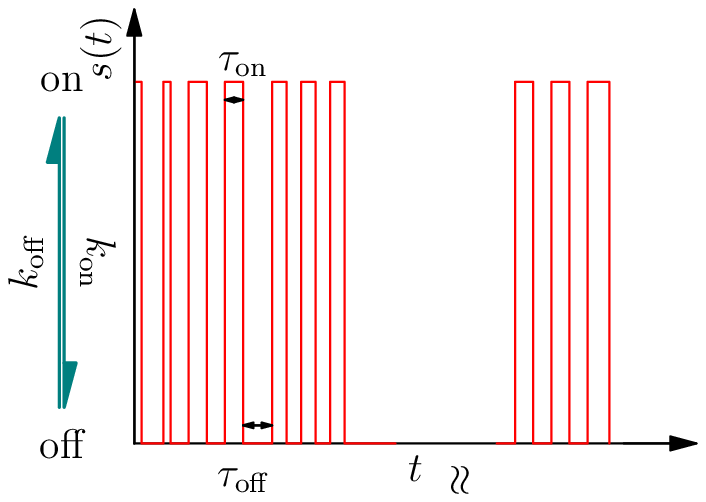}
    \label{Figure1d}}
  \caption{\small (Color online) Schematic of the two-well model for the MSC
    gating. (a) Energy profile with prestress $\gamma_0$ and (b) with
    additional prestress $\gamma=\gamma_0+\Delta \gamma$ (b); $A$ is the MSC
    in-plane surface area and $\Delta A$ is its change when opening up;
    $\Delta h$ is the intrinsic energy barrier in the absence of applied
    tension, $\ka$ is the area stretch modulus of the harmonic profiles, taken
    identical for both wells. (c) and (d) Associated time series $s(t)$ of the
    residence periods $\cT$ spent in open or closed states.}
  \label{Figure1}
\end{figure}
% ---------------------------------------

%
Mechanosensitive ion channels (MSCs) are present in nearly all cell
types~\cite{Martinac}; they are integral membrane proteins responding over a
wide dynamic range to mechanostimuli subsequently transduced into
electrochemical signals~\cite{Arnadottir}. There appear to be two modes of
action for MSCs: (i) those that receive stress from fibrillar proteins
resulting in gating, and (ii) cases in which tension in the surrounding
bilayer forces the channel to open. Our focus is on the latter type---the
stretch-activated channels---in which the stimulus mechanically deforms the
membrane's lipid bilayer that, in turn, triggers MSC conformational changes
through an intricate mechanical coupling~\cite{Arnadottir,Ursell}. It is
important to recall that the high sensitivity of the cellular mechanosensory
apparatus does not originate from the MSCs themselves but from an efficient
coupling between the channel gating machinery and the cellular structures that
transmit the force~\cite{Orr}. The existence of calcium-based
stretch-activated MSCs in the amoeba \textit{Dictyostelium discoideum} has
recently been revealed by Lombardi \textit{et al.}~\cite{Lombardi}, which is
believed to be at the root of its shear-flow induced motility~\cite{Decave}
improved by calcium mobilization~\cite{Fache}.

MSCs adopt conformational states with distinct functional properties in
response to the applied tension along the plane of the cell membrane instead
of the normal pressure~\cite{Gustin,Sokabe&Sachs,Sokabe}. The gating of these
transient receptor channels is to a good approximation represented by a
two-state double-well model~\cite{Ursell,Sukharev&Corey}~[see
Figs.~\ref{Figure1a} and~\ref{Figure1b}].  Directional mechanosensing requires
cells to make accurate decisions based on biased stochastic transitions
between MSC conformational states [see Figs.~\ref{Figure1c}
and~\ref{Figure1d}]. Although a fundamental bound on the accuracy of
directional chemical gradient sensing was derived~\cite{Endres,HuSNR}, no
theory exists for the physical limits of directional mechanosensing.

\section{Single mechanosensitive ion channel sensing}
% --------------------------------------------------------------------------

In this section, our focus is the physical limitations in sampling by a single
MSC, subject to a mild shear flow inducing minute changes $\Delta \gamma\ll
\gamma_0$ to the lateral membrane tension $\gamma = \gamma_0+\Delta \gamma$.

\subsection{Two-state double-well model for the gating mechanism}
% -----------------------------------------------------------------------------------------
%
We consider a single MSC, which is a specialized transmembrane protein that
can undergo a distortion in response to external mechanical forces applied
through the lipid bilayer itself. At its simplest, this mechanical deformation
can be described as a conformational transition between closed and open states
separated by a free energy barrier denoted as $\Delta h$. In the particular
case of the gating of the well-studied bacterial large conductance
mechanosensitive channel MscL, the energy difference between the closed and
the fully open states in the unstressed membrane was found to be $18.6\kb T$
with an associated energy barrier $\Delta h \sim 38\kb
T$~\cite{Sukharev}. Without loss of generality, we assume that both
conformational states, open and closed, are symmetrically positioned with
respect to the free energy barrier, which implies that the absolute area
change between the bottom of each wells is $\Delta A/2$. We account for the
elasticity of each state---assumed identical and harmonic for both states---by
considering a quadratic dependence of the free energy in the lateral membrane
tension $\gamma$~\cite{Sukharev&Corey}. The unidirectional transition rates,
given in Eyring's form, are
\begin{align}
  \kon&=k_0\exp\left(-\frac{\gamma\Delta A/2}{\kb T}-\frac{\gamma^2A}{2\ka\kb
      T}\right),\\
  \koff&=k_0\exp\left(\frac{\gamma \Delta A/2-\Delta h}{\kb
      T}-\frac{\gamma^2A}{2\ka\kb T}\right),
\end{align}
where $k_0$ is a scaling factor, $A$ is the MSC in-plane surface area, $\Delta
A$ is its change in the in-plane area when opening up, $\gamma$ is the lateral
membrane tension, and $K_A$ is the area stretch modulus. For clarity, we omit
the thermal energy $\kb T$ in what follows. We consider, here, a weak
mechanostimulus inducing minute changes $\Delta \gamma$ in the membrane
tension $\gamma=\gamma_0+\Delta \gamma$ with $\Delta \gamma \ll \gamma_0$,
$\gamma_0$ being the cell's membrane prestress. At the first order in $\Delta
\gamma$, the unidirectional transition rates can be expressed as
\begin{align}
  \kon&=k_0\exp\left(-\frac{\gamma_0\Delta
      A}{2}-\frac{\gamma_0^2A}{2\ka}\right)\exp\left( -(1+\alpha) \frac{q}{2}  \right),\\
  \koff&=k_0\exp\left(\frac{\gamma_0\Delta A}{2}-\Delta
    h-\frac{\gamma_0^2A}{2\ka}\right)\exp\left( (1-\alpha) \frac{q}{2}
  \right),
\end{align}
where $q=\Delta \gamma \Delta A$ is the extra work generated by the
extracellular mechanical signals. The nondimensional parameter $\alpha=2
\gamma_0A/(\ka \Delta A)$ represents the ratio of the total energy
$\gamma_0\Delta A/2$ expanded for the in-plane deformation of the MSC to the
energy $\ka (\Delta A/2)^2/A$, associated with the membrane thinning due to
the membrane volume conversation~\cite{Ursell}. In the particular case of MscL
gating, one finds $\alpha \sim 0.58$, given that $\gamma_0=3.5\kb
T/\textrm{nm}^2$, $\Delta A=6\textrm{~nm}^2$, $A=30\textrm{~nm}^2$ and
$\ka=60\kb T/\textrm{nm}^2$~\cite{Ursell,Rawicz,Sukharev}.

Such a perturbation $\Delta \gamma$ to the lateral membrane tension induces a
stretching of the MSC, triggering its opening if the associated free energy
surpasses the barrier $\Delta h$. An internal feedback mechanism is
responsible for closing down the MSCs which are relentlessly switching between
open and closed states (see Fig.~\ref{Figure1}). This dynamics is
characterized by the binary sequence $s(t)$, spent in both possible
states. This process is essentially a Markovian telegraph process: Memoryless
transitions are entirely determined by a switching
rate~\cite{Gillespie}. Therefore the lengths of open and closed intervals have
exponential distributions with means $1/\kon$ and $1/\koff$, respectively,
$\kon$ and $\koff$ being the unidirectional transition rates in conformational
states.

\subsection{Signal estimation by linear regression}
% -------------------------------------

%
To know how well a cell can determine the shear stress applied to its
membrane, it is assumed that information is derived from its MSC states based
on the concept of ``perfect instrument'' registering switching
events~\cite{Berg&Purcell}. MSCs switch between open and closed states with
$s(t)=1$ for $t \in \ton $ and $s(t)=0$ for $t \in \toff $. We use the time
series $s(t)$---as being the time record of MSC states measured by a perfect
instrument---to investigate the dynamics of a given MSC over a long signal
exposure time, i.e. for $\cT \gg 1/\kon$ and $\cT \gg 1/\koff$. We perform a
linear regression (LR) of the binary time series $s(t)$.  In the limit of long
time series $\cT$, with starting time $t_0$, the mean and variance of $s(t)$
over the of observation are classically given by~\cite{Gillespie},
\begin{align}
  S&= \frac{1}{\cT}\int_{t_0}^{t_0+\cT} s(t)\,\text{d}t = \frac{\kon}{\kon+\koff},\\
  \sigma_s^2&=\langle (\delta s)^2 \rangle = \frac{\kon\koff}{(\kon+\koff)^2}.
\end{align}
Still, in the limit of long time series,
\begin{equation}
  S \simeq \<s\>=\frac{\kon}{\kon+\koff}\bigg|_{q=\tilde{q} }=\frac{1}{1+\exp(\tilde{q}-\Delta \heff)},
\end{equation}
where $\<s\>$ is the ensemble average of $s(t)$, $\tilde{q}$ is the true value
of $q$, and $\Delta \heff=\Delta h-\gamma_0\Delta A$ is the effective free
energy barrier reduced by the existing prestress action. The signal can be
inferred from the fraction of MSC active time $S$ with $S \simeq \langle s
\rangle$ for long $\cT$. To compute the variance of $S$, the covariance of
$s(t)$ is needed, and it can be calculated directly from its definition,
\begin{align}
  G(t,t')\equiv\langle s(t)s(t')\rangle - {\langle s(t)\rangle}^2 & =
  \sigma_s^2 e^{-|t-t'|/\tau},\nonumber \\
  &= \sigma_s^2 e^{-|t-t'|(\kon+\koff)}.
\end{align}
If we repeat this observation many times, starting at wildly different times
$t_0$, the variance of $S$ is
\begin{equation}
  \sigma_S^2=\frac{1}{{\cal T}^2}\int_{0}^{{\cal T}}\text{d}t\int_{0}^{{\cal T}}\text{d}t' G(t,t')=\frac{2}{\cT}\frac{\koff\kon}{(\koff+\kon)^3}\label{eq:variance}.
\end{equation}
A standard LR yields
\begin{equation}
  \delta q=\frac{\kon}{\koff}\delta \left( \frac{S}{1-S}\right)=\frac{\kon}{\koff}\frac{\delta S}{S^2}\label{q0},
\end{equation}
and the following estimate for $q=\Delta \gamma \Delta A$:
\begin{equation}
  q^{\rm LR}=\Delta \heff+\ln \frac{S}{1-S}=\Delta \heff+\ln \frac{\ton}{\toff}\label{q0_1},
\end{equation}
where $\Delta \heff$=$\Delta h - \gamma_0 \Delta A$ is the effective energy
barrier, reduced by the existing membrane prestress $\gamma_0$.  From
Eqs.~\eqref{q0} and \eqref{q0_1}, we obtain the associated variance,
\begin{equation}
  \sigma_q^2=\frac{\kon^2}{\koff^2} \frac{\sigma_S^2}{S^4}=\frac{2(\kon+\koff)}{\cT(\kon\koff)}=\frac{2 }{n},\label{eq:sqlr}
\end{equation}
in terms of the number of registered switches $n$ defined as
\begin{equation}
  n\equiv \cT \frac{\kon\koff}{\kon+\koff},
\end{equation}
and physically representing the number of transitions between the two
conformational states. Note that if $\kon \ll \koff$ or $\kon \gg \koff$, $n$
can simply be expressed as $n\simeq \cT/\textrm{max}(\kon^{-1},\koff^{-1})$.

\subsection{Signal estimation using a maximum likelihood
  estimator} \label{appMLE}
% ---------------------------------------------------------------------------
%
It is still unclear how exactly a cell performs its signal estimation based on
the register of switching events. The LR presented in the previous section
appears as the most rudimentary form of statistical estimation. Alternatively,
a maximum likelihood estimate (MLE)~\cite{Kay} can be sought for the two-state
discrete-valued telegraph process which is generated by switching values at
jump times of a Poisson process~\cite{Gillespie}.

For a long exposure to a signal---i.e., for large $\cT=\ton+\toff$--- and
given the unidirectional transition rates $\kon$ and $\koff$, the likelihood
function is obtained by acknowledging the fact that we are in the presence of
a stationary Poisson process,
\begin{equation}
  \cL=\frac{(\kon \ton)^n}{n!}  e^{-\kon
    \ton}\cdot\frac{(\koff \toff)^{n}}{n!} e^{-\koff \toff} ,\label{eq:eq1}
\end{equation}
where $\ton$ (respectively, $\toff$) is the total open (respectively, closed)
time and $n$ is the number of switching events. When omitting the unessential
constant terms, the log-likelihood function is cast as
\begin{equation}
  \ln \cL=-(\kon \ton+\koff \toff)+n\ln(\koff\kon).
\end{equation}
The MLE is considered to provide an estimate of $q$. To this aim, maxima of
the first-order derivative of the above log-likelihood are sought
\begin{align}
  \left(\frac{\partial \ln \cL }{\partial q}\right)(q=\qmle)&= -\kon \ton
  (1+\alpha)\nonumber \\
  -&\koff \toff (1-\alpha)+2 n\alpha=0,
\end{align}
which yields
\begin{equation}
  \qmle =\Delta \heff+\ln\frac{\ton}{\toff}.
\end{equation}
To quantify the uncertainty associated with the above maximum likelihood
estimation, one has to consider the second-order derivative of the
log-likelihood function,
\begin{equation}
  \frac{\partial^2 \ln {\cal L}}{\partial q^2}=-\kon \ton (1+\alpha)^2-\koff \toff (1-\alpha)^2
\end{equation}
to ascertain the normalized variance in the long exposure to the signal limit,
\begin{equation}\label{eq:sqmle}
  \sigma_q^2=-\left<\left(\frac{\partial^2 \ln \cL}{\partial q^2}\right)(q=\qmle)\right >^{-1}=\frac{2 }{(1+\alpha^2)n}.
\end{equation}
According to the Cram\'er-Rao lower bound (CRLB), the variance $\sigma_q^2$
sets the lowest measurement uncertainty through sampling~\cite{Kay}.

The uncertainties of mechanosensing using LR and MLE [Eqs.~\eqref{eq:sqlr}
and~\eqref{eq:sqmle}] show that for a given stimulus exposure, statistical
fluctuations limit the precision with which a single MSC can determine the
stimulus amplitude. Similar to chemosensing, MLE yields a more accurate
mechanosensing lower limit than LR~\cite{Endres}, albeit for fundamentally
different reasons. Indeed, the two estimates for $q$ given by the LR and the
MLE are identical, whereas, the associated variances are different. In this
particular problem, the linear regression is intrinsically limited by its
linear character and only captures the lowest-order term which does not
involve $\alpha$. This is, of course, no longer the case with the MLE. From a
physical standpoint, it is like the LR is not able to account for the thinning
effects of the lipid bilayer; the variations in the thickness of the lipid
bilayer are negligible for high values of $\ka$, i.e. for near-zero values of
$\alpha$. The very presence of $\alpha=2 \gamma_0A/(\ka \Delta A)$ in
Eq.~\eqref{eq:sqmle} highlights the connection between the mechanical
properties of the cell and the measurement uncertainty~\cite{zeroalpha}.

\section{Limits of cellular mechanosensing}
% ----------------------------------------------------------

In this section, our focus is the physical limitations in sampling by an array
of MSCs distributed across the cell, subject to a mild shear flow inducing
nonuniform minute changes $\Delta \gamma\ll \gamma_0$ to the lateral membrane
tension $\gamma = \gamma_0+\Delta \gamma$.
\subsection{Model of a cell subjected to a linear shear flow}
% ------------------------------------------------------------------------------

%
We now turn to directional mechanosensing by an entire cell, focusing on the
idealized case of $N$ uniformly distributed MSCs on the equator (only) of a
spherical cell of radius $R$. Observing the MSC distribution is experimentally
challenging, but it is very unlikely that it is homogeneous. For the sake of
analytical simplicity, our model does not consider this fact. We assume the
MSCs to be independent, neglecting inter-MSC interactions.  One might argue
that local interactions among the MSCs could result globally in a cooperative
effect which may help smaller cells better discriminate the signal
direction---see Ref.~\cite{Duke} regarding the cooperativity between chemical
receptors for chemotactic \textit{E. coli}. The present analysis would, thus,
provide conservative estimates for this problem.

\begin{figure}[htbp]
  \includegraphics[clip,width=.3\textwidth]{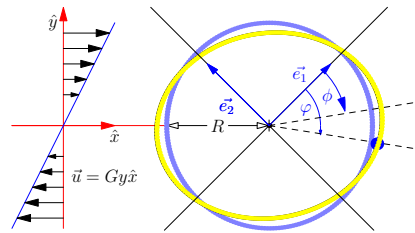}
  \caption{\small (Color online) Cell deformation under a linear shear flow
    with shear rate $G$. The elliptic curve represents the intersection of the
    ellipsoid with the $xy$ plane. The circular curve is the initial membrane,
    and $\vec{e}_1$ is the direction of the largest elongation rate
    eigenvector~\cite{Marmottant}; the dot represents a given MSC.}
  \label{Figure2}
\end{figure}

Fluid shear stress, which occurs naturally in a variety of physiological
conditions, is one of the most important
mechanostimuli~\cite{Arnadottir,ShearFlow,Park,Moares}. Furthermore, cell
locomotion generates Stokes flows which can be sensed by neighboring
cells~\cite{Bouffanais,Moares}. Specifically, fluid shear stress induces a
nonuniform tension on the cell's lipid bilayer triggering an asymmetric
stretch activation of some MSCs, themselves giving rise to an intracellular
biochemical cascade driving pseudopod extensions preferentially in the
direction of the tension gradient~\cite{ShearFlow}. At the cell's microscale,
any natural flow field approximates locally to a linear shear flow (see
Fig.~\ref{Figure2}).  For an artificial spherical cell (vesicle) subject to
small deformations due to a weak mechanical stimulus, the tension distribution
at the equator (see Fig.~\ref{Figure2}) reads~\cite{Marmottant}
\begin{align}\label{eq:gamma}
  \gamma(\varphi) &= \gamma_0 + \Delta \gamma (\varphi),\\
  \Delta \gamma & = - \frac{5}{4}\eta G R \cos 2 (\varphi-\phi),
\end{align}
$\eta$ being the viscosity and $\phi$ being the phase angle difference between
the minimum tension point and the largest elongation axis~\cite{Marmottant}.
An MSC located in a high-tension zone has a higher probability to open
up. This spatial asymmetry creates an angular bias in the fluctuations of the
$N$ time traces $\mathbf{S}=\{S_1,\ldots,S_N\}$ across the cell, $S_i$ being
the fraction of open state of the $i$th MSC at location $\varphi=
\varphi_i$. We prove that, by a global statistical processing of $\mathbf{S}$,
a cell can infer the stimulus direction. The uncertainty due to the ubiquitous
and limiting presence of noise is also derived.

\subsection{Statistics for the shear-stress induced signal at the cellular
  level}
% ----------------------------------------------------------------------------------------------

%
When exposing a cell to shear stress [see Fig.~\ref{Figure2} and
Eq.~\eqref{eq:gamma}], the nonuniform perturbation in its membrane tension
$\Delta \gamma$ induces an uneven MSC redistribution across the cell. Using
the white-noise approximation, the conformational state of the $i$-th MSC at
$\varphi_i$, subject to $q_i=\Delta \gamma(\varphi_i) \Delta A= -Q \cos
2(\varphi_i-\phi)$ with $Q=\frac{5}{4}\eta G R \Delta A$, is $ S_i = \langle
S_i \rangle + \eta_i$ with
\begin{equation}
  \langle S_i \rangle  =
  \frac{\kon^i}{\kon^i+\koff^i}=\frac{1}{1+\exp(-\Delta \heff-Q \cos
    2(\varphi_i-\phi) ),}
\end{equation}
and
\begin{equation}
  \langle \eta_i \eta_j\rangle  =
  \frac{2}{\cT}\frac{\koff^i\kon^i}{(\koff^i+\kon^i)^3}\delta_{ij}=
  \sigma_{S_i}^2 \delta_{ij},
\end{equation}
where $\sigma_{S_i}^2$ takes the form of Eq.~\eqref{eq:variance} at the $i$th
location. The MSC signal $\mathbf{S}$ is a vector of independent Gaussian
random variables with different means but approximately identical variances
$\sigma_{S}^2$. From Eq.~\eqref{eq:variance}, we find that $\sigma_S^2$
decreases as $\cT$ increases, with $\sigma_S^2\rightarrow 0$ in the limit of
$\cT\rightarrow \infty$.  Instead of time averaging over long exposure to
signal time $\cT$, we consider ensemble averaging over $m$ independent MSCs
subject to the same signal, thus giving $S = \frac{1}{m}\sum_{k=1}^{m}
s_k$. The variance associated with this ensemble averaging is
\begin{equation}
  \sigma_S^2=\frac{1}{m}\frac{\kon\koff}{(\kon+\koff)^2}\label{EA}.
\end{equation}
From Eqs.~\eqref{eq:variance} and~\eqref{EA}, one can establish that a single
MSC observed over time $\cT$ is statistically equivalent to ensemble averaging
over $m\equiv\frac{1}{2}\cT(\kon+\koff)=\frac{\cT}{2\tau}$ independent
MSCs. This allows us to recast the white Gaussian noise component as
\begin{equation}
  \langle \eta_i \eta_j\rangle =
  \frac{1}{m}\frac{\koff^i\kon^i}{(\koff^i+\kon^i)^2}\delta_{ij}.
\end{equation}
As we are working in the limit of small membrane deformations induced by a
mild mechanical stimulus, we expand $\langle S_i\rangle$ in small
$Q=\frac{5}{4}\eta GR\Delta A$ up to the leading order
\begin{align}
  \langle S_i \rangle&\simeq \langle S \rangle -\mu \cos
  2(\varphi_i-\phi),\\
  \intertext{with} \langle S \rangle &= \frac{\kon}{\kon+\koff}\bigg|_{Q=0},
\end{align}
and where $\mu=m\sigma_S^2 Q$ is the signal amplitude. At the first order in
$Q$ for $S_i$, we also have
\begin{equation}
  \langle \eta_i \eta_j\rangle \simeq
  \frac{1}{m}\frac{\koff\kon}{(\koff+\kon)^2}\bigg|_{Q=0}\delta_{ij}=\sigma_S^2(Q=0)
  \delta_{ij}.
\end{equation}
To summarize, at the leading order in $Q$,
\begin{equation}\label{eq:signal}
  S_i\approx \langle S \rangle -m\sigma_S^2 Q\cos2(\varphi_i-\phi)+\eta_i,
\end{equation}
where $\{\langle S\rangle,\sigma_S^2\}
=\left\{\frac{\kon}{\kon+\koff},\frac{\kon\koff}{m(\kon+\koff)^2}\right\}_{|Q=0}$. The
associated signal-to-noise ratio (SNR)~\cite{Kay} is
\begin{equation}
  \kappa\equiv \frac{\mu^2}{\sigma_S^2}=m^2 {\sigma_S^2} Q^2.
\end{equation}

\subsection{Maximum likelihood estimation of the magnitude and direction of
  the mechanostimulus}
% ---------------------------------------------------------------------------------------------------------

%
The signal~\eqref{eq:signal} has a classical form---sinusoidal in phase with
added white Gaussian noise---commonly encountered in signal processing
applications~\cite{Kay}. Estimating the shear-flow direction for the cell is
strictly equivalent to estimating the phase $\phi$ of~\eqref{eq:signal}. Given
the nonlinear nature of the relationship between the mechanostimulus and the
spatiotemporal signal available to the cell, a nonlinear statistical
estimation is required.  A nonlinear MLE of $\btheta=\{Q,\phi\}$ can be
achieved by resorting to the jointly sufficient statistics~\cite{Kay} given
by:
\begin{align}
  z_1&=\sum_{i=1}^N(S_i-\langle S\rangle)\cos2\varphi_i,\\
  z_2&=\sum_{i=1}^N(S_i-\langle S\rangle)\sin2\varphi_i.
\end{align}
The associated joint probability density function reads
\begin{align}
  p(\mathbf{Z})=\frac{2}{\pi \sigma_S^2 N}\exp&\left( -\frac{N
      \mu^2}{4\sigma_S^2}+\frac{\mu}{\sigma_S^2}(z_1\cos2\phi-z_2\sin2\phi)\right)\nonumber \\
  \times &\exp\left(-\frac{1}{N\sigma_S^2}(z_1^2+z_2^2)\right),
\end{align}
leading to the following expression:
\begin{equation}
  p(\mathbf{Z})=\frac{2e^{-N\kappa/4}}{\pi \sigma_S^2 N}
  \exp\left[\frac{\mu (z_1\cos2\phi-z_2\sin2\phi)}{\sigma_S^2}-\frac{z_1^2+z_2^2}{N\sigma_S^2}\right].
\end{equation}
Thus, the likelihood function, $\cL=p(\mathbf{Z}|\mathbf{\Theta})$, is given
by the joint probability density function which gives access to the
log-likelihood
\begin{align}
  \ln \cL &=-\frac{N \mu^2}{4\sigma_S^2}+\frac{\mu}{\sigma_S^2}(z_1\cos2\phi-z_2\sin2\phi)-\frac{1}{N\sigma_S^2}(z_1^2+z_2^2),\nonumber\\
  &= m Q(z_1\cos2\phi-z_2\sin2\phi) -\frac{m^2N\sigma_S^2}{4}Q^2.\
\end{align}
The maximum likelihood estimators for the vector parameter
$\mathbf{\Theta}=\{Q,\phi\}$ are defined to be the value that maximizes the
likelihood function over that allowable domain for $\mathbf{\Theta}$ and it is
found from
\begin{align}
  \frac{\partial \ln p(\mathbf{Z}|\mathbf{\Theta})}{\partial Q}(Q=\Qmle)&=0,\\
  \frac{\partial \ln p(\mathbf{Z}|\mathbf{\Theta})}{\partial
    \phi}(\phi=\phimle)&=0
\end{align}
yielding, respectively,
\begin{align}
  \Qmle&=\frac{2\sqrt{z_1^2+z_2^2}}{N m\sigma_S^2 },\\
  \phimle&=-\frac{1}{2}\arctan\frac{z_2}{z_1}.
\end{align}
The CRLB yields the respective variances of the MLE,
$\hat{\btheta}_{\textrm{\tiny{MLE}}}=(\Qmle,\phimle)$, which are calculated by
means of the inverse of the Fisher information matrix, which is the negative
of the expected value of the Hessian matrix~\cite{Kay},
\begin{align}
  \sigma_{\hat{Q}}^2 &= -\left<\frac{\partial^2 \ln {\cal L}}{\partial
      Q^2}\right >^{-1}=\frac{2 }{ N m^2
    \sigma_S^2},\label{eq:varianceQ}\\
  \sigma_{\hat{\phi}}^2 &= -\left<\frac{\partial^2 \ln {\cal L}}{\partial
      \phi^2}\right >^{-1}=\frac{1 }{2 N m^2 \sigma_S^2 \hat{Q}^2}=\frac{1}{2
    N{\kappa}},\label{eq:variancePhi}
\end{align}
where $\kappa\equiv \mu^2/\sigma_S^2$ is the SNR.  Both uncertainties in the
signal amplitude and phase are inversely proportional to $m=\frac{\cT}{2\tau}$
and $\sqrt{N}$, i.e., favoring longer signal exposure time $\cT$ with as many
MSCs as possible.

\subsection{Analysis of the results}
% -----------------------------------------------

%
Typically, eukaryotic cells are 10--100 $\unit{\upmu m}$ across with uniform
MSC surface density on the order of $1/\unit{\upmu m^2}$~\cite{Morris,Davies},
and the number of MSCs increases with the cell surface area $A$ as $N=N_0
R^2$. Given that $Q\sim R$, we get the paramount fact that the SNR $\kappa$
varies like $R^4$; this amounts to an enormous $10^4$ ratio in SNRs for small
and large eukaryotic cells. Larger cells are considerably more effective at
directional mechanosensing. Interestingly, we also find that $\kappa\propto
1/{\sigma_{\hat{\phi}}^{2}}$, exactly like the case of eukaryotic directional
gradient chemosensing, despite fundamental differences in signaling
mechanisms~\cite{HuSNR}.

\begin{figure}[htbp]
  \subfigure[] {
    \includegraphics[width=4cm]{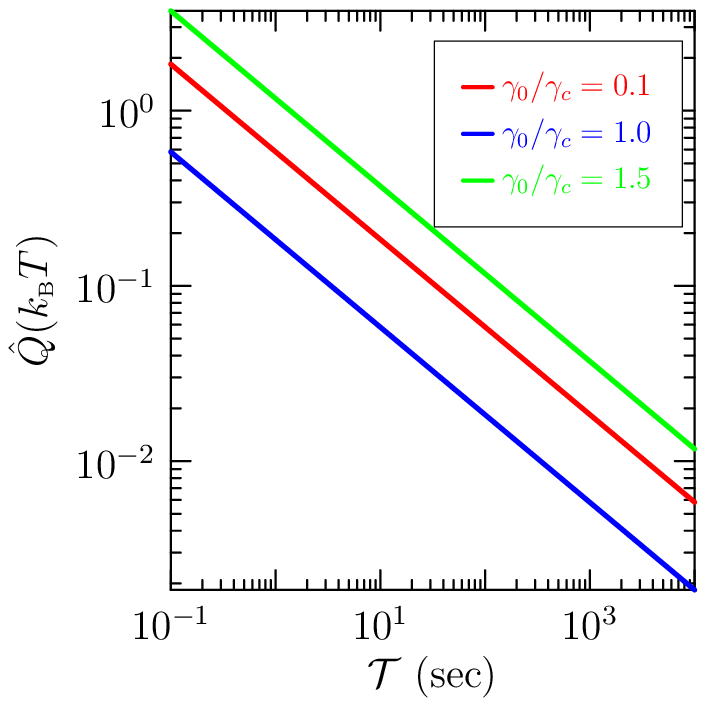}
    \label{fig:3a}} \subfigure[] {
    \includegraphics[width=4cm]{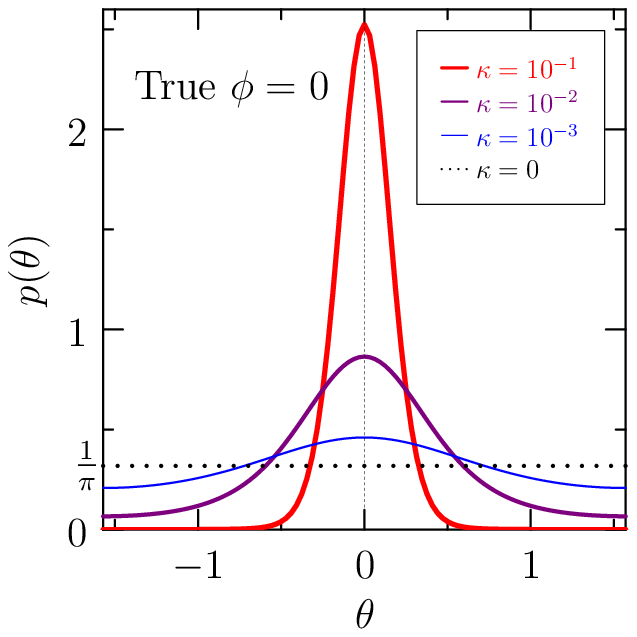}
    \label{fig:3b}}
  \caption{(Color online) (a) Relationship between observation time $\cT$ and
    MLE estimated signal amplitude $\hat{Q}$, for $\gamma_0/\gamma_c=0.1,\ 1,\
    1.5$; (b) probability density function (PDF) of the phase estimate for a
    true value $\phi=0$ for different SNR values $\kappa$. The following
    values are used~\cite{Ursell,Sukharev&Corey}: $\ka=60\kb T/\unit{nm}^2$,
    $\Delta h=38 \kb T$, $A=30 \unit{nm}^2$, $\Delta A=10 \unit{nm}^2$,
    $\gamma_c=1.0\kb T/\unit{nm}^2$, $1/k_0=1 \unit{m s}$, and $N=200$.}
  \label{Figure4}
\end{figure}

Understanding how the SNR relates to cell characteristics---specifically
elastic and gating properties---allows one to uncover some essential features
of eukaryotic directional mechanosensing. For a given $\cT$, the SNR reads
\begin{equation}\label{eq:snrfinal}
  \kappa=n \hat{Q}^2  =\left[\frac{k_0\cT}{2}\frac{\exp(-\gamma_0^2A/(2\ka)-\gamma_0\Delta
      A/2)}{1+\exp(\Delta h -\gamma_0\Delta A )}\right] \hat{Q}^2,
\end{equation}
$n$ being the number of switching events. Considering varying prestresses
$\gamma_0$ in Eq.~\eqref{eq:snrfinal}, one finds that $\kappa$ achieves large
values about its maximum attained at the critical prestress $\gamma_c=\Delta
h/\Delta A$ such that $\Delta \heff=0$. At this point, it is worth
highlighting that the cortical cytoskeleton structurally supports the fluid
bilayer, thus, providing the cell membrane with a shear rigidity that is
lacking in simple bilayer vesicles~\cite{Hamill}. Through membrane
fluctuations and membrane trafficking, the cell has the ability to regulate
and to tune the prestress of its lipid bilayer~\cite{Hamill,Davies}---see
Ref.~\cite{Rivero} for more details on the case of \textit{Dictyostelium}
cells. Note that one limitation of our model is the lack of information with
regard to the energy barrier $\Delta h$, which prevents us from explicitly
finding the value of the critical prestress $\gamma_c=\Delta h/\Delta A$.

It is revealing to study the relationship~\eqref{eq:variancePhi} between
$\hat{Q}$ and exposure time $\cT$ required by the cell to detect the stimulus
direction for different prestress values [see Fig.~\ref{fig:3a}]. Cells with a
near-critical prestress require a much shorter exposure to the signal. It
would be interesting to experimentally measure, for various types of cells,
the critical prestress and to compare it to the actual prestress. According to
our results, this experiment should reveal a significantly much higher
mechanosensitivity of cells such that $\gamma_0\simeq \gamma_c$. Strikingly,
the scale of $\hat{Q}$ can be as small as $ 10^{-2}\kb T$ for a cell with
$200$ MSCs exposed over $\cT \sim 10^3$~s. This fact is clearly related to the
growing evidence of exquisite sensitivity of cells to
mechanostimuli~\cite{Arnadottir,Moares,Davies}. In addition, cells having one
or both of the characteristics of near-critical prestress $\gamma_0\simeq
\gamma_c$, and low gating energy barrier $\Delta h$, will benefit from a
higher SNR, resulting in improved directional mechanosensing capabilities. On
the other hand, cells not satisfying one of the above conditions or subjected
to a higher background noise, might see their SNR falling below an estimation
threshold point SNR $\kappa^*$---the point at which the cell is no longer able
to estimate the stimulus direction. Indeed, this estimation process is
essentially nonlinear---owing to the nonlinear relationship between the
mechanostimulus and the spatial signals registered by the cell---and thereby
suffers from a low SNR threshold effect induced by the appearance of outlying
peaks in the log-likelihood function~\cite{Kay,Richmond}. Here, an MLE is
considered, but any other type of statistical estimation that exhibits such a
nonlinear threshold effect constitutes a serious fundamental limit in the
cell's ability to effectively perform directional mechanosensing at a low
SNR~\cite{Kay}. It is important noting that the very existence of this
estimation threshold is only contingent upon the nonlinear nature of the
relationship between the stimulus and the spatiotemporal signal processed by
the cell. No general analytical expression for the estimation threshold point
SNR $\kappa^*$ exists, even in the particular case of the nonlinear MLE
considered here. However, Monte Carlo simulations could be considered to
numerically estimate $\kappa^*$ for any given nonlinear statistical estimation
techniques, including the MLE.

\subsection{Specifics of high signal-to-noise ratios cellular mechanosensing}
% ----------------------------------------------------------------------------------------------------

%
At the other extreme, for large SNRs, MLE is asymptotically unbiased,
efficient, and delivers a fine prediction of the uncertainty in the
mechanostimulus direction [see Eq.~\eqref{eq:variancePhi}].  Expressing
$p(\mathbf{Z})$ using polar coordinates with $(z_1,z_2)=(\rho\cos
2\theta,-\rho \sin 2\theta)$, where the latter minus sign is introduced to
obtain a symmetric PDF,
\begin{equation}
  p(\rho,\theta)=\frac{2}{\pi \sigma_S^2 N}e^{-N\kappa/4}\exp\left( \frac{\mu\rho}{\sigma_S^2}\cos(2\phi-2\theta)-\frac{\rho^2}{N\sigma_S^2}\right).
\end{equation}
Hence, the symmetric kernel function is given by
\begin{align}
  p(\theta)&=\int_0^{\infty}p(\rho,\theta)\textrm{d}\rho=\frac{2}{\pi
    \sigma_S^2 N}e^{-N\kappa/4}\nonumber \\
  \times &\int_0^{\infty}\rho\exp\left(
    \frac{\mu\rho}{\sigma_S^2}\cos(2\phi-2\theta)-\frac{\rho^2}{N\sigma_S^2}\right)\textrm{d}\rho,
\end{align}
and the PDF of the phase $\phi$ estimate reads
\begin{equation}\label{eq:PDF}
  p(\theta)=\frac{e^{-N\kappa/4}}{\pi}\left[1+{\sqrt \pi b(\theta)
      e^{b^2(\theta)}}\left(1+\operatorname{erf}\left(b(\theta)\right)\right)\right],
\end{equation}
where erf is the canonical error function and
\begin{equation}
  b(\theta)= \frac{\sqrt{N\kappa}}{2}\cos 2(\phi
  -\theta).
\end{equation}
We now consider the case of a high SNR $\kappa$, for which the phase estimate
will be near its true value. Therefore, using the approximation
$\cos2(\phi-\theta)\simeq 1$ and the identity $\cos^2(x)=1-\sin^2(x)$ yields
\begin{align}
  & p(\theta)\simeq\frac{1}{\pi}\exp(-N\kappa/4)\\
  &+\sqrt{\frac{ {N\kappa} }{4\pi
    }}\exp\left[-\frac{N\kappa}{4}\sin^22(\phi-\theta)\right]\left[1+\operatorname{erf}\left[\frac{\sqrt{N\kappa}}{2}\right]\right].\nonumber
\end{align}
For high SNRs, the first term in the above equation and the error function in
the second term will be approximately 1. In the limit $\kappa \rightarrow
\infty$, this PDF tends asymptotically to a classical Gaussian PDF given by
\begin{equation}\label{eq:PDFGaussian}
  p(\theta)\simeq\sqrt \frac{N\kappa}{\pi }\exp\left(-N\kappa (\phi-\theta)^2\right),
\end{equation}
for which the variance is directly accessible:
\begin{equation}
  \sigma_{\phi}^2=\frac{1}{2
    N\kappa},
\end{equation}
and is found to be identical to the variance $\sigma_{\hat{\phi}}^2$ [see
Eq.~\eqref{eq:variancePhi}] obtained using the CRLB for $\phimle$. Thus, the
dependence $\sigma_{\hat{\phi}}^2 \propto 1/\kappa$ is asymptotically
recovered and holds for $\kappa > \kappa^*$. For $\kappa<\kappa^*$,
$\sigma_{\hat{\phi}}^2$ rises sharply until a so-called no information point
is reached~\cite{Richmond}. The no information region corresponds to very low
SNRs, i.e,. $\kappa \rightarrow 0$, where the PDF is nearly uniform
$p(\theta)\simeq1/\pi$, thus, preventing the cell from extracting any
directional information from $\mathbf{S}$. As already mentioned in the
previous section, a closed-form expression of $\kappa^*$ is yet to be found
for this nonlinear estimation problem. However, a value for $\kappa^*$ and its
asymptotic relationship with the uncertainty in directional mechanosensing
could be established experimentally or computationally. The above discussion
is well illustrated by looking at the PDF of the phase estimate [see
Eq.~\eqref{eq:PDF}] for widely different SNRs shown in Fig.~\ref{fig:3b}: At a
high SNR $\kappa=10^{-1}$, the PDF is almost Gaussian which is consistent with
both the MLE results (estimator and variance) and the asymptotic
expression~\eqref{eq:PDFGaussian}. For an intermediate SNR $\kappa=10^{-2}$,
the PDF deviates from its asymptotic Gaussian form, whereas, the MLE deviates
from the CRLB. For even lower SNRs, $\kappa=10^{-3}$ and $\kappa=0$, the cell
has passed the estimation threshold point and has entered the no information
region. It should be added that the maximum value and the tail of the
PDF~\eqref{eq:PDF} for varying SNRs are vastly different from those of the
Gaussian PDF~\eqref{eq:PDFGaussian}.

\section{Conclusions}
% ---------------------------

Despite its relative simplicity, our biophysical model sheds some light on the
physical limits of cellular directional mechanosensing, which prove to exhibit
many similarities with its chemical counterpart: higher accuracy for large
cells and $\sigma_{\hat{\phi}}^2\propto 1/\kappa$.

More specifically, we found that the signal-to-noise ratio varies like $R^4$,
where $R$ is a measure of the cell's size. Experimentally, this could easily
be verified by considering two types of amoebae of typical sizes approximately
10 $\upmu$m and $100$ $\upmu$m respectively, and by subjecting them to the
same mild mechanostimulus in the same environment, i.e., with the same
background noise.

This model also reveals how the biochemical nature of the cell's membrane
impacts cellular directional mechanosensing. Indeed, we showed the existence
of a critical prestress which entirely depends on the free energy barrier---
this energy barrier is fixed for one particular type of MSC. Therefore, for
one particular type of cell, if the prestress value for the lipid bilayer
happens to be close to the critical prestress, then the mechanosensitive
process benefits from a much higher signal-to-noise ratio. This could be
tested experimentally with various different types of cells, having notably
different natures of their lipid bilayers and, hence, different prestress
values. This set of cells would have to be subjected to the same
mechanostimulus of decreasing magnitude under the same environmental
conditions.

Finally, we uncovered the existence of another fundamental limit in the
cellular directional mechanosensing owing to the nonlinear nature of the
relationship between the mechanostimulus and the spatial signals registered by
the cell. Indeed, all nonlinear statistical estimation techniques, including
the one used by the cell, intrinsically suffer from the appearance of a low
SNR threshold effect beyond which the signal estimation can no longer be
considered as reliable.


\begin{thebibliography}{reference}
  % ----------------------------------
%
\bibitem{Arnadottir} J.~\'Arnad\'ottir and M.~Chalfie,
  Annu. Rev. Biophys.~\textbf{39}, 111~(2010).~C.~Kung, B.~Martinac and
  S. Sukharev, Annu. Rev. Microbiol.~\textbf{64}, 313~(2010).
%
\bibitem{ShearFlow} E.~D\'ecav\'e, D.~Rieu, J.~Dalous, S.~Fache, Y.~Br\'echet,
  B.~Fourcade, M.~Sartre and F.~Brucket, J.~Cell~Sci.~\textbf{116},
  4331~(2003); A.~Makino, E.~R.~Prossnitz, M.~B\"unemann, J.~M.~Wang,
  W.~Yao. and G.~W.~Schmid-Sch\"onbein,~Am.~J.~Cell~Physiol.~\textbf{290},
  C1633~(2006).
%
\bibitem{Moares} C.~Moares, Y.~Sun and C.~A.~Simmons,
  Integr. Biol.~\textbf{3}, 959~(2011).
%
\bibitem{Park} J.~Y.~Park, S.~J.~Yoo, L.~Patel, S.~H.~Lee, S.~H.~Lee,
  Biorheology~\textbf{47}, 165~(2010).
%
\bibitem{Olesen} S.~P.~Olesen, D.~E.~Clapham and P.~F.~Davies,
  Nature~\textbf{331}, 168 (1988); E.~C.~Jacobs, C.~Cheliakine,
  D.~Gebremedhin, P.~F.~Davies and D.~R.~Harder, FASEB J.~\textbf{7}, 71
  (1993).
\bibitem{Davies} P.~F.~Davies, Physiol. Reviews~\textbf{75}, 519 (1995).
%
\bibitem{Decave} E.~D\'ecav\'e, D.~Rieu, J.~Dalous, S.~Fache, Y.~Br\'echet,
  B.~Fourcade, M.~Sartre and F.~Bruckert, J. Cell Sci.~\textbf{116}, 4331
  (2003).
%
\bibitem{Orr} A.~W.~Orr, B.~P.~Helmke, B.~R.~Blackman and M.~A.~Schwartz,
  Developmental Cell~\textbf{10}, 11 (2006).
%
\bibitem{S&S} S.~Sukharev and F.~Sachs, J. Cell Sci.~\textbf{125}, 3075
  (2012).
%
\bibitem{Rawicz} W.~Rawicz, K.~C.~Olbrich, T.~McIntosh, D.~Needham and
  E.~Evans, Biophys. J.~\textbf{79}, 328 (2000).
%
\bibitem{Opsahl} L.~R.~Opsahl and W.~W.~Webb, Biophys. J.~\textbf{66}, 75
  (1994); C.~E.~Morris and U.~Homann U, J. Membr. Biol.~\textbf{179}, 79
  (2001); V.~S.~Markin and F.~Sachs, Phys. Biol.~\textbf{1}, 110 (2004).
%
\bibitem{Bouffanais} R.~Bouffanais and D.~K.~P.~Yue, Phys. Rev. E~\textbf{81},
  041920 (2010).
%
\bibitem{Martinac} B.~Martinac and A.~Kloda,
  Prog. Biophys. Mol. Biol.~\textbf{82}, 11 (2003).
%
\bibitem{Ursell} T.~Ursell, J.~Kondev, D.~Reeves, P.~A.~Wiggins and
  R.~Phillips, The role of lipid bilayer mechanics in mechanosensation. In
  Mechanosensitive Ion Channels. A.~Kamkin and I.~Kiseleva (eds.),
  Springer-Verlag, Berlin. Chap. 2, pp.~37--70, (2008).
%
\bibitem{Lombardi} M.~L.~Lombardi, D.~A.~Knecht and J.~Lee, Exp. Cell
  Res.~\textbf{314}, 1850 (2008).
%
\bibitem{Fache} S.~Fache, J.~Dalous, M.~Engelund, C.~Hansen, F.~Chamaraux,
  B.~Fourcade, M.~Sartre, P.~Devreotes and F.~Bruckert, J. Cell
  Sci.~\textbf{118}, 3445 (2005).
%
\bibitem{Gustin} M. C. Gustin, X.~L.~Zhou, B.~Martinac and C.~Kung, Science
  \textbf{242}, 762 (1988).
%
\bibitem{Sokabe&Sachs} M. Sokabe and F. Sachs, J. Cell Biol.  \textbf{111},
  599 (1990).
%
\bibitem{Sokabe} M. Sokabe, F.~Sachs and Z.~Q.~Jing, Biophys. J.
  \textbf{599}, 722 (1991).
%
\bibitem{Sukharev&Corey} S. Sukharev and D.~P. Corey, Sci. STKE \textbf{2004},
  re4 (2004).
%
\bibitem{Endres} R.~G.~Endres and N.~S.~Wingreen,
  Phys. Rev. Lett.~\textbf{103}, 158101~(2009); T.~Mora and N.~S.~Wingreen,
  Phys. Rev. Lett.~\textbf{104}, 248101~(2010).
%
\bibitem{HuSNR} B. Hu, W.~Chen, W.-J.~Rappel and H.~Levine,
  Phys. Rev. Lett. \textbf{105}, 048104 (2010); B. Hu, W.~Chen, H.~Levine and
  W.-J.~Rappel, J. Stat. Phys. \textbf{142}, 1167 (2011).
%
\bibitem{Sukharev} S~I. Sukharev, W.~J. Sigurdson, C.~Kung and F. Sachs,
  J. Gen. Physiol.~\textbf{113}, 525, (1999).
%
\bibitem{Gillespie} D.~T.~Gillespie,~\textit{Markov Processes: An Introduction
    For Physical Scientists} (Academic Press, San Diego, CA, 1992), Chap.~6.
%
\bibitem{Berg&Purcell} H.~C.~Berg and E.~M.~Purcell, Biophys~J.~\textbf{20},
  193 (1977).
%
\bibitem{Kay} S.~M.~Kay, \textit{Fundamentals of Statistical Signal
    Processing: Estimation Theory} (Prentice Hall, Upper Saddle River, NJ,
  1993), Vol.~1.
%
\bibitem{zeroalpha} For $\alpha=0$, both estimators yield the same estimate
  and accuracy; the LR cannot capture the quadratic dependence of free energy
  on tension.
%
\bibitem{Duke} T.~A.~J.~Duke and D.~Bray,
  Proc. Natl. Acad. Sci. USA~\textbf{96}, 10104 (1999).
%
\bibitem{Marmottant} P.~Marmottant, T.~Biben and S.~Hilgenfeldt,
  Proc. R. Soc. London A~\textbf{464}, 1781 (2008).
%
\bibitem{Morris} C.~E.~Morris, J.~Membrane~Biol.~\textbf{113}, 93 (1990).
%
\bibitem{Hamill} O.~P.~Hamill and B.~Martinac, Physiol. Reviews~\textbf{81},
  685 (2001).
%
\bibitem{Rivero} F.~Rivero, B.~Koppel, B.~Peracino, S.~Bozzaro, F.~Siegert,
  C.~J.~Weijer, M.~Schleicher, R.~Albrecht and A.~A.~Noegel, J. Cell
  Sci.~\textbf{109}, 2679 (1996).
%
\bibitem{Richmond} C.~D.~Richmond, IEEE Trans. Inf. Theory~\textbf{52}, 2146
  (2006).
%
\end{thebibliography}
\end{document}